\documentclass[a4paper,american,british,english,3p,times,twocolumn]{elsarticle}
\usepackage[T1]{fontenc}
\usepackage[latin9]{inputenc}
\usepackage{amsbsy}
\usepackage{graphicx}
\PassOptionsToPackage{normalem}{ulem}
\usepackage{ulem}

\makeatletter

\usepackage{ecrc}

\volume{00}

\firstpage{1}

\journalname{Nuclear and Particle Physics Proceedings}

\runauth{Malkov}

\jid{nppp}

\jnltitlelogo{Nuclear and Particle Physics Proceedings}
\journal{Example: Nuclear Physics B}

\usepackage{hyperref}

\DeclareMathAlphabet\mathbfcal{OMS}{cmsy}{b}{n}

\makeatother

\usepackage{babel}
\begin{document}

\begin{frontmatter}{}
\selectlanguage{british}%

\title{Newly-Discovered Anomalies in Galactic Cosmic Rays: Time for Exotic
Scenarios?}

\author{M.A. Malkov}

\address{University of California, San Diego, La Jolla, CA 92093}
\selectlanguage{english}%
\begin{abstract}
Recent observations of galactic cosmic rays (CR) in the 1-500 GeV
energy range have revealed striking deviations from what deemed ``standard.''
The anomalies cut across hadronic and leptonic CRs. I discuss findings
that challenge physical mechanisms long held responsible for the CR
production in galactic supernova remnants (SNR). I also consider some
new physics of particle acceleration in SNR shocks that is not part
of conventional models but may explain the anomalies. However, a possible
20-30\% excess remains unaccounted for in the $e^{+}/e^{+}$ ratio
over the range of a few 100 GeV. If not explained by future models,
it suggests an additional source of positrons such as a dark matter
decay/annihilation or pulsar contribution. Earlier efforts to explain
both the $e^{+}/e^{-}$ and $p$/He anomalies with the ``standard''
models by adjusting the SNR environmental parameters and multiple
sources are critically assessed. 
\end{abstract}
\begin{keyword}
cosmic rays\sep supernova remnants \sep particle acceleration \sep
dark matter \PACS 98.38.Mz \sep 98.70.Sa \sep 96.50.S- \sep 95.35.+d 
\end{keyword}

\end{frontmatter}{}

\section{Recent Challenges}

In the two years passed after the first ``Cosmic Ray Origin\textendash Beyond
the Standard Model'' San Vito Conference \cite{BlandfordCRorig2014},
new discoveries moved the CR modeling farther beyond all ``standards''
than perhaps over the preceding decade, which was also rich in surprises.
Some models explaining the data were jettisoned after the next data
release or a careful analysis of those available \cite{VladimirMoskPamela11}.
The latest debates seem to refocus on\foreignlanguage{british}{ a
perplexing \emph{rise of the }$e^{+}/e^{-}$\emph{ fraction} in the
10-300 GeV range, and possibly related flatness of $\bar{p}/p$ and
$e^{+}/p$ ratios. They gained widespread interest after the AMS-02
team has published the most accurate elemental and energy spectra
in this range \cite{AMS02_antipro2016}. }

\selectlanguage{british}%
Anomalies in the CR spectra and composition are becoming a general
trend in CR observations. Besides the $e^{+}/e^{-}$ anomaly, there
is at least one interesting phenomenon that merits a short discussion,
as it may also help to understand the $e^{+}$ excess.

\subsection{Proton to Helium, Carbon, and Oxygen ratios}

In 2011 Pamela experiment \cite{Adriani11} revealed a remarkable
$0.1$ index difference between the proton and He rigidity spectra
in 2-200 GV range. The limits here are contingent rather than intrinsic
to the original spectrum, merely set by the solar modulation and instrument's
range. The scaling is likely to continue beyond this range, according
to high-energy observations \cite{CREAM11}. 

With this finding, the Pamela team caught the Holy Grail of the CR
origin hypothesis, the diffusive shock acceleration (DSA) mechanism
in the cross-hair. It is widely believed to supply the bulk of galactic
CR from their presumed production sites in supernova remnant (SNR)
shocks. \foreignlanguage{english}{To appreciate the threat to the
DSA, consider equations of motion for a particle in electromagnetic
fields $\mathbf{E}\left(\mathbf{r},t\right)$ and $\mathbf{B}\left(\mathbf{r},t\right)$,
using the particle rigidity $\mathbfcal R=\mathbf{p}c/eZ$ instead
of its momentum $\boldsymbol{\mathbf{p}}$ ($Z$ is the charge number),
e.g., \cite{MDSPamela12}: }

\selectlanguage{english}%
\begin{equation}
\frac{1}{c}\frac{d\mathbfcal R}{dt}=\mathbf{E}\left(\mathbf{r},t\right)+\frac{\mathbfcal R\times\mathbf{B}\left(\mathbf{r},t\right)}{\sqrt{\mathcal{R}_{0}^{2}+\mathcal{R}^{2}}},\label{eq:RigMotion}
\end{equation}

\begin{equation}
\frac{1}{c}\frac{d\mathbf{r}}{dt}=\frac{\mathbfcal R}{\sqrt{\mathcal{R}_{0}^{2}+\mathcal{R}^{2}}}.\label{eq:coordMotion}
\end{equation}
The fields $\mathbf{E}\left(\mathbf{r},t\right)$ and $\mathbf{B}\left(\mathbf{r},t\right)$
are arbitrary, so our inference below will apply to both the acceleration
and propagation of the CRs. It follows that all species with $\mathcal{R}\gg\mathcal{R}_{0}=Am_{p}c^{2}/Ze$
($A$ is the atomic number and $m_{p}$- proton mass, so $\mathcal{R}_{0}\sim A/Z$
GV), have identical orbits in the phase space $\left({\bf r},\mathbfcal R\right)$. 

Let us select, at the moment $t=0$, a group of H$^{+}$ and He$^{++}$
ions undergoing shock acceleration from the same $\mathcal{R}\left(0\right)\gg\mathcal{R}_{0}$.
Let their ratio at $t=0$ be $\eta=N_{p}/N_{{\rm He}}$. According
to eqs.(\ref{eq:RigMotion}) and (\ref{eq:coordMotion}), it is conserved
during the \emph{acceleration and propagation}. Physically, $\eta$
may depend on $\mathcal{R}\left(0\right),$ only if the strong inequality
$\mathcal{R}\left(0\right)\gg\mathcal{R}_{0}$ is violated. But, as
long as $\eta\left(t=0\right)$ is fixed and the acceleration is stationary,
we can assign $\eta$ to a different $\mathcal{R}=\mathcal{R}_{1}\gg\mathcal{R}_{0}$
beyond which the above argument about conservation of $\eta$ applies.

The individual values of $N_{p}\left(\mathcal{R}\right)$ and $N_{{\rm He}}\left(\mathcal{R}\right)$
will decline with growing \emph{$\mathcal{R}$} since particles gradually
escape the accelerator and then galaxy. But the \emph{ratio} $N_{p}/N_{{\rm He}}$
must remain constant because the escape mechanisms are the same for
$p$ and He. As we know, $N_{p}/N_{{\rm He}}$ does depend on $\mathcal{R},$
nonetheless. Two effects can be responsible for that: spallation and
time-variability of injection. We will use the rigidity dependence
of the \emph{fractions }of different species as a primary probe into
the intrinsic properties of CR accelerators. Unlike the individual
spectra, the fractions are unaffected by the CR propagation, reacceleration,
and losses from the galaxy.

Meanwhile, the spallation effects can hardly account for the $p$/He
rigidity dependence \cite{VladimirMoskPamela11}. Furthermore, the
rigidity independence of acceleration and propagation rules out ``easy
solutions.'' The 0.1 $p$/He index cannot be caused by propagation
effects, second-order Fermi (re)acceleration, or compression-expansion
cyclic acceleration in the ISM (interstellar medium) turbulence, contrary
to several statements made in the literature along these lines. Accepting
the obvious, let us leave alone scenarios that are inconsistent with
the dynamical equivalence of $p$ and He ions at $\mathcal{R}\gg\mathcal{R}_{0}$.

\selectlanguage{british}%
More recently, also the $p/C$ and $p/O$ ratios have been measured
to have the same rigidity scaling as  $p/$He, that is $\propto\mathcal{R}^{-0.1}$
\footnote{\selectlanguage{english}%
http://www.ams02.org/wp-content/uploads/2016/12/Final.pdf\selectlanguage{british}%
}. Constant C/O and He/C ratios are expected from eqs.(\ref{eq:RigMotion})
and (\ref{eq:coordMotion}), provided that these elements are injected
in a fully ionized state (see Sec.\ref{subsec:Shock's-intrinsic-mechanism}
though). Therefore, He, C, and O are likely to share their acceleration
and propagation history, so it is unlikely that C and O are pre-accelerated
from grains \cite{Ohira2016PhRvD}. Moreover, the equivalence between
the He, C and O spectra corroborates an earlier conclusion \cite{VladimirMoskPamela11}
that spallation effects are insufficient to account for the observed
differences in rigidity spectra between $p$ and other elements with
similar $A/Z$. 

\subsection{Positron anomaly\label{subsec:Positron-anomaly}}

\selectlanguage{english}%
The condition $\mathcal{R}\gg\mathcal{R}_{0}$ for the equivalence
between $p$ and He is irrelevant for $e^{+}$ and $e^{-}$, particles
of equal mass, whereas the charge sign now matters most. If $e^{+}$and
$e^{-}$ are injected into the same or similar shocks, their spectral
differences \foreignlanguage{british}{(by contrast to the secondary
positrons}\footnote{\selectlanguage{british}%
Note that the injected positrons can be seeded as secondaries, which
we assume to be the case and discuss in Sec.\ref{sec:ModelingPositrons}.\selectlanguage{english}%
}\foreignlanguage{british}{) must come from \emph{charge-sign} sensitive
acceleration and/or propagation. Most mechanisms for injection, acceleration,
and propagation rely on particle interactions with MHD waves, self-driven
by the particles or preexisting. However, such interactions do not
seem to offer any distinct charge-sign dependence. }

\selectlanguage{british}%
Much can be learned about wave-particle interactions from a linear
resonance condition $\omega-k_{\parallel}v_{\parallel}=n\omega_{c}.$
Here $\omega$ and $k_{\parallel}$ are the frequency and wave-vector
projection on the averaged magnetic field direction, $v_{\parallel}$
is that of the particle velocity. The particle charge sign enters
$\omega_{c}=eB_{0}/\gamma mc$, but the integer $n$ can be positive/negative
(for normal/anomalous Doppler resonance), or zero (for Cerenkov resonance).
In the latter case the interaction is charge-sign independent, so
we focus on the cyclotron resonances, $n=\pm1$. Consider an individual,
circularly polarized wave propagating along the field line. In its
own frame, where the particle speed is $v_{\mathcal{k}}^{\prime}=v_{\mathcal{k}}-\omega/k_{\mathcal{k}}$,
$\mathbf{E}=0$ and the particle perceives only the Lorenz force.
The resonance occurs when this force oscillates in sync with the particle
gyromotion, $k_{\mathcal{k}}v_{\mathcal{k}}^{\prime}=\pm\omega_{c}$.
The choice of the sign ($\pm$) depends on the wave polarization while,
physically, the condition means that the particle perceives a non-oscillating
force (strong effect). 

Consequently, once protons resonantly generate waves of a particular
polarization, these waves will resonantly interact with positrons
with the same $v_{\mathcal{k}}^{\prime}/\omega_{c}$ and, also with
electrons moving at $-v_{\mathcal{k}}^{\prime}$. These considerations
are not strictly correct if the wave grows to make the resonance condition
nonlinear \cite{m98}. Such a situation occurs near a collisionless
shock transition to which we return later. Here we note that under
typical DSA conditions when $v_{\mathcal{k}}\gg U_{{\rm shock}}$,
particles are isotropic in pitch angle. The isotropy implies a nearly
equal amount of them with positive and negative $v_{\mathcal{k}}^{\prime},$
so it is hard to expect a significant suppression of electron acceleration
compared to the positrons.

\selectlanguage{english}%
\begin{figure}
\includegraphics[bb=0bp 0bp 500bp 170, scale=0.76]{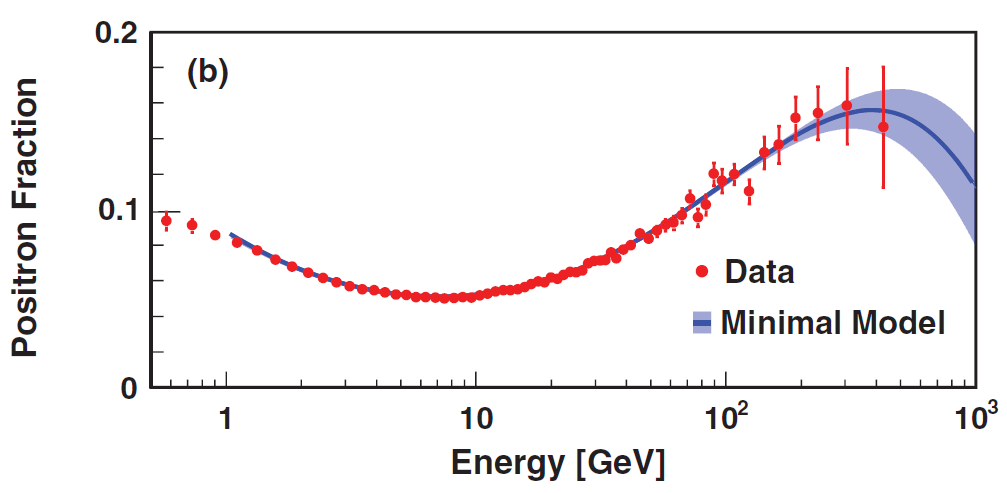}
\includegraphics[bb=0bp 0bp 500bp 380, scale=0.48]{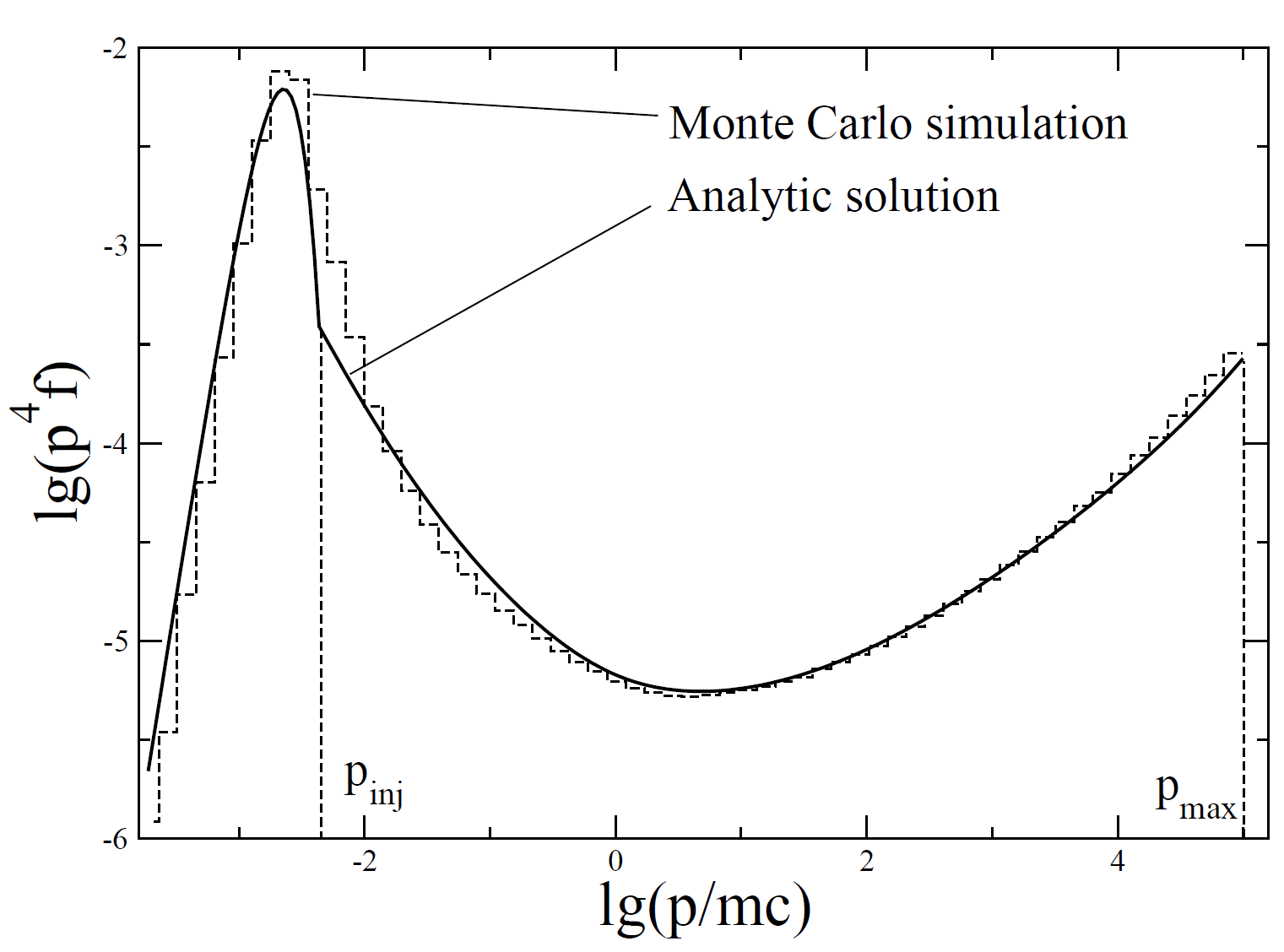}

\caption{Top Panel: $e^{+}/\left(e^{+}+e^{-}\right)$ fraction measured by
AMS-02 \cite{AMS02_2014}.\protect \\
Bottom Panel: Analytic solution obtained in \cite{m97a} for NLDSA
of protons and adopted from \cite{Mosk07} (shown together with Monte
Simulations from \cite{EllisBerBar00}). \label{fig:PosFracAndNLDSAanalSol}}
\end{figure}

\selectlanguage{british}%
However plausible this conclusion may seem, the $e^{+}/\left(e^{+}+e^{-}\right)$
ratio strongly depends on energy \cite{Pamela_Pos_09,AMS02_Pos_13}
(Fig.\ref{fig:PosFracAndNLDSAanalSol}, top panel). Moreover, by contrast
with the He/\foreignlanguage{english}{$p$} ratio that shows a featureless
scaling $\propto\mathcal{R}^{0.1}$ (at least in the $\mathcal{R}>\mathcal{R}_{0}$
range) the $e^{+}/\left(e^{+}+e^{-}\right)$ ratio possesses one or
possibly even two extrema. Strong growth at high energies attracted
the most attention, but a distinct minimum at $\approx8$ GeV and
descending branch in the range $E<8$ GeV may equally be relevant.
This nonmonotonic positron fraction is hard to explain with a minimum
of assumptions. It appears particularly at odds with a \emph{single
source} DSA operation in an SNR, which predicts similar rigidity spectra
for all species accelerated in the source. The data indicate that
this ratio has a trend towards saturation at 200-400 GeV, or may even
decline beyond this energy range. So, a maximum in this range is also
possible which, of course, has not passed unnoticed by the dark matter
hunters.

\section{Possible Physics behind the anomalies}

\selectlanguage{english}%
Both $e^{+}$ and He spectral anomalies violate the equal rigidity
rule, whereby two particles are dynamically indistinguishable if they
have equal rigidities, regardless of mass or charge. In the first
example, the difference between $p$ and He was in their rest-mass
rigidity, namely $\mathcal{R}_{0,{\rm He}}=2\mathcal{R}_{0,p}$ ($\mathcal{R}_{0,{\rm He}}=\mathcal{R}_{0,{\rm C}}=\mathcal{R}_{0,{\rm O}}$,
for fully ionized ions). It explicitly enters the equations of motion
but becomes unimportant when particles are accelerated to $\mathcal{R}\gg\mathcal{R}_{0}$.
From this moment on their rigidity spectra (apart from normalization)
must be the same, which they are not. In the second example, the absolute
value of rigidity was the same but the signs were opposite ($e^{\pm}$).
Accepting the arguments for charge-sign symmetry of wave-particle
interaction, the most likely mechanism behind the $e^{\pm}$ separation
during or before the acceleration phase is a macroscopic electric
field. Indeed, an accelerator-scale electric field can turn positrons
toward it while fending off the electrons, or vice versa. A legitimate
concern about a plasma quasineutrality in such field is not a dogma
but rather a problem in plasma physics. 

\begin{figure}
\selectlanguage{english}%
\includegraphics[bb=0bp 0bp 500bp 300bp, scale=0.6]{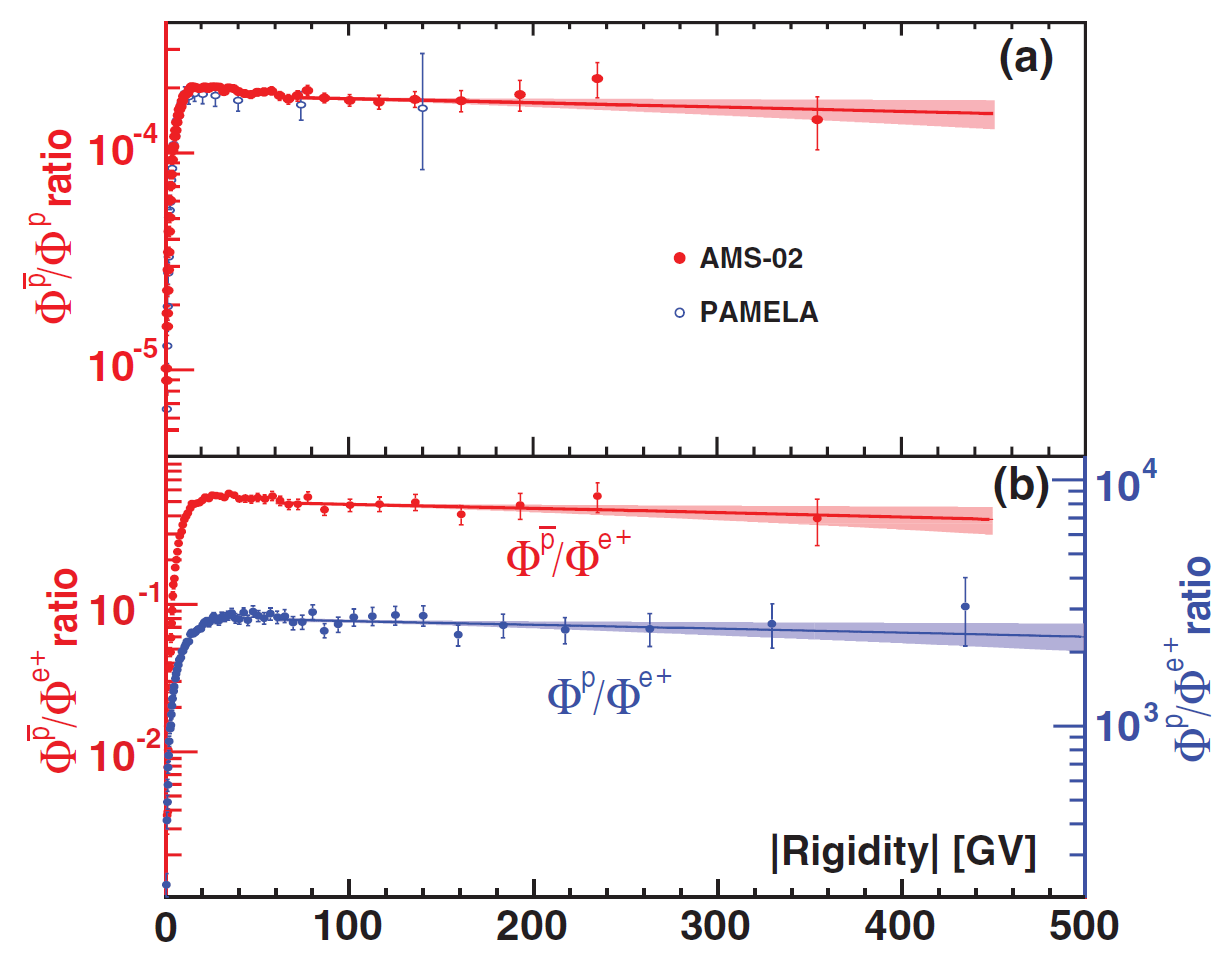}

\selectlanguage{british}%
\caption{$\bar{p}/p$ and $e^{+}/p$ fractions measured by AMS-02 and Pamela.
\label{fig:AntiprotonsAMS}}
\end{figure}

\selectlanguage{british}%
The positron excess also hints at an exciting but still speculative
involvement of weakly interacting massive particles (WIMPs). The WIMPS
are widely believed to be present in the dark sector of the particle
theories beyond the standard model \cite{DMtheory2009PhRvD,Berezinsky2015JPhCS}.
The search for them in CR data is argued to be in a synergistic relation
with that in the LHC. Decay or annihilation of such particles has
been suggested as a possible explanation for the rise in the $e^{+}/e^{-}$
ratio, as well as surprisingly flat (for the secondaries) $\bar{p}/p$
ratio, Fig.\ref{fig:AntiprotonsAMS}.

\selectlanguage{english}%
Let us lay aside the WIMP scenarios for a while. Earlier considerations
give enough clues as to where should one look for the explanations.
In the case of the p/He anomaly, the particle segregation must occur
during the non-relativistic phase of acceleration. Ideally, it must
be part of the shock injection mechanism at $\mathcal{R}\ll\mathcal{R}_{0}$.
This would be the way to produce relatively more He than protons with
growing rigidity. However, this alone will not solve the problem,
if the shock accelerates particles under stationary conditions, first
of all at a constant Mach number. Fortunately, the Mach number of
an SNR shock decreases in time, and the mechanism may work. Of course,
it requires the relative injection efficiency of p/He to be also Mach
number dependent which is true. 

\selectlanguage{british}%

\selectlanguage{english}%
In the case of the $e^{+}/e^{-}$ anomaly, a macroscopic electric
field can separate the species very efficiently. Below, we consider
$p/$He and $e^{+}/e^{-}$ spectral anomalies independently and compare
the physical mechanisms outlined above with environmental mechanisms,
suggested in the literature.

\section{Modeling of $p$/He Spectrum}

The rigidity-dependent $p$/He ratio seems paradoxical from the perspective
of a single accelerator but makes more sense if the $p$ and He spectra
come from two or more sources. 

\subsection{Environmental and multisource scenarios}

Multisource models (see \cite{Serpico15} for a recent review) were
suggested before the high-quality measurements \cite{Adriani11,AMS02He2015PhRvL}
have become available. The models were adjusted for the newest data
\cite{Tomassetti2015ApJ}, but their intrinsic limitations in accounting
for the persistent $0.1$ difference between the He and $p$ indices
remain and can be readily assessed. For instance, if only two sources
with fixed power-law indices and normalization for H and He contribute
to the observed spectrum, they cannot generate a nearly constant slope
of $N_{p}/N_{{\rm He}}\propto\mathcal{R}^{-0.1}$ over two decades
in rigidity (at least!). This slope can occur only in a transition
region between two regions each of which corresponds to one particular
source with a constant slope. By adding more sources and adjusting
propagation parameters \cite{Tomassetti2015ApJ} one can extend the
transition region somewhat. However, unless the sources with the hardest
spectra are proton deficient, it is hard to imagine that He and heavier
elements approach the observed proton flux in the knee region \cite{Kascade2014}.
They must maintain a consistently harder spectrum relative to protons
for all $\mathcal{R}\gg\mathcal{R}_{0}$ which is impossible if only
one stationary source dominates the knee region.

\selectlanguage{british}%
A different scenario, which may operate at a single SNR exploding
into a superbubble, assumes an inhomogeneity of an SNR shock environment
with an outward increasing H/He ratio \cite{Ohira2016PhRvD}. The
idea is that as the shock expands into such an environment, the ratio
of the processed H/He ions also increases. Since the shock weakens
with growing radius, the integrated H spectrum will be softer than
that of He. However, to generate a $0.1$ difference in the spectral
indices, the radial dependence of the He abundance must be coordinated
with the shock Mach number $M$ dependence on its radius, $M\left(R\right)$,
since the power-law index $q=4/\left(1-M^{-2}\right)$ is the same
for all species. While not impossible for He, the same radial profile
of C and O abundance seems improbable. Recall, that all three elements
have been measured to have identical spectra. Besides, the explosion
must occur near the maximum of the He, C, and O abundances. Had it
occurred in the region where the abundance changes monotonically,
the effect would not be pronounced. The latter concern may be resolved
by assuming a layered pre-supernova wind structure \cite{Bier2010},
but again, He, C, and O need to have proper radial profiles. 

\subsection{Shock's intrinsic mechanism \label{subsec:Shock's-intrinsic-mechanism}}

The above mechanisms are contingent in that they depend on how the
CR sources are distributed in the ISM, types of SN progenitor stars,
or their environments. But can a shock change an element abundance
by selectively accelerating it out of a homogeneous background plasma?
It has been argued for some time \cite{m98} that precisely this occurs
in quasi-parallel shocks. What is relevant to the present discussion,
the selectivity mechanism significantly depends on the shock Mach
number.

At an intuitive level, the main argument is of a Le Chatelier's type.
Injected protons distort the shock structure in such a way as to reduce
the injection. Quasi-parallel shocks thus promote the CR diversity
by \foreignlanguage{american}{disfavoring} acceleration of the most
abundant species. The segregation mechanism works as follows. Injected
protons drive unstable Alfven waves in front of the shock. These waves
control the injection of all particles by regulating their access
to those parts of the phase space from where they can cross and re-cross
the shock. As protons drive these waves, the waves trap protons most
efficiently. Furthermore, the waves are almost frozen into the local
fluid so, when crossing the shock interface, they entrain most particles
and prevent them from escaping upstream, thus significantly reducing
their odds for injection. Again, most efficient is namely the proton
entrainment, while, e.g., alpha particles have somewhat better chances
to escape upstream and to get eventually injected. The wave trapping
of alphas is weaker because of the mismatched wavelengths generated
by the protons since their mass to charge ratios are different by
a factor of two. The trapping becomes naturally stronger with growing
wave amplitude, and this trend is more pronounced for the protons.
This difference is the crux of the injection selectivity.

\begin{figure}
\selectlanguage{english}%
\includegraphics[bb=0bp 0bp 500bp 240bp,scale=0.70]{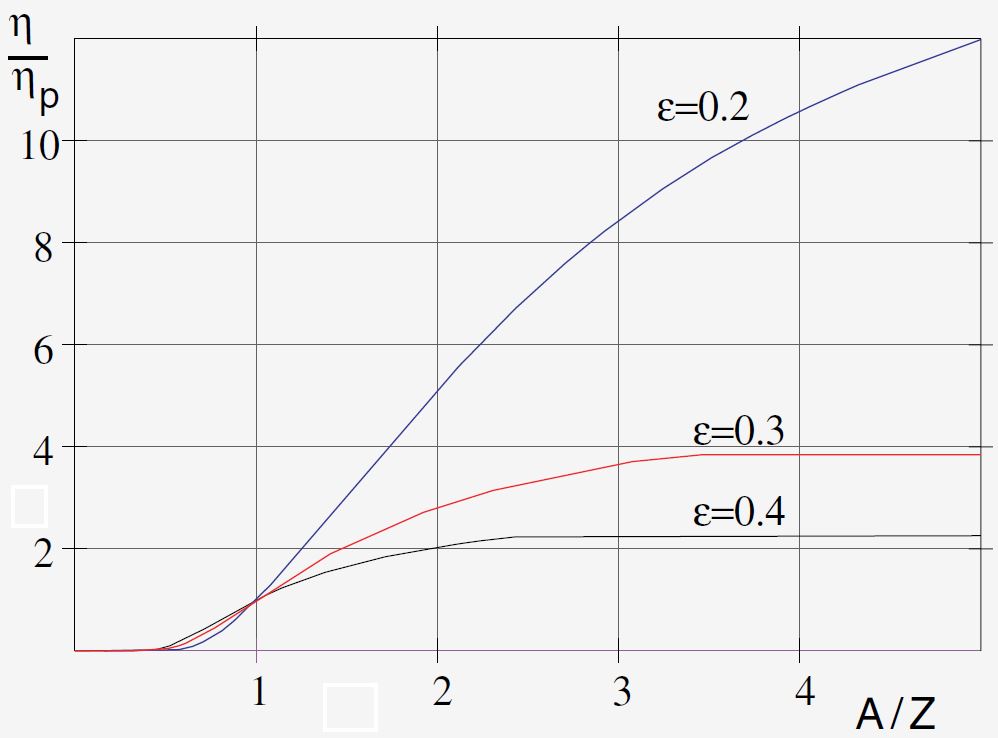}

\selectlanguage{british}%
\caption{Elemental injection efficiencies normalized to proton efficiency depending
on mass to charge ratio A/Z, shown for different amplitude parameters
$\epsilon=B_{0}/\delta B_{2}$, where $\delta B_{2}\simeq r\delta B_{1}$
is the shock-compressed (compression ratio $r$) wave amplitude. For
strong shocks $\epsilon\sim1/3$ and it slowly decreases with the
Mach number (adopted from \cite{m98}).\label{fig:Elemental-injection-efficiencies}}
\end{figure}

It follows that the injection rate must depend on the mass to charge
ratio, $A/Z$. The charge-sign dependence is less significant, so
we discuss it later. The injection efficiency is shown in Fig.\ref{fig:Elemental-injection-efficiencies}
as a function of $A/Z$ for a few values of the wave amplitude parameter
$\epsilon=B_{0}/\delta B<1$ downstream. The latter quantity is thus
determined by the amplitude of the wave generated by upstream protons
undergoing injection (shock reflected or leaking from downstream).
If the elements under consideration are fully ionized before injection,
He, C, and O are injected at the same rate since they have similar
$\mathcal{R}_{0}$. However, even if C, for example, is not fully
ionized, thus having a larger $A/Z$, it will not make much difference,
as the injection curves saturate rather quickly. Therefore, an incomplete
ionization will likely produce only an insignificant increase of the
carbon injection. Moreover, when the shock Mach number is high, the
assumption about a fully ionized plasma appears reasonable. Also,
the validity range of the curves shown in Fig.\ref{fig:Elemental-injection-efficiencies}
is limited from above. If $A/Z$ increases significantly, such species
becomes unaffected by the proton-driven waves and their injection
will drop (not just saturated, as shown) because they will remain
downstream after the first shock passing. Therefore, injection of
grains with $A/Z\gg1$ is unlikely. An \emph{ad hoc }assumption about
the presence of long waves generated by CRs at a knee rigidity ($\gtrsim10^{14}$GV)
is needed to thermalize such heavy grains and get them injected \cite{Ellison1997}.
It is clear that waves generated upstream by injected protons are
much more powerful than those long waves required to scatter grains,
so we focus on the short waves. 

The presence of such strong waves, significantly amplified upon shock
crossing is well established by observations and simulations \cite{Quest88}.
However, it is not clear what is the primary source of the wave driving
protons. One physically motivated source of such particles is an \emph{evaporation}
of hot downstream plasma into the upstream medium, as suggested by
Parker \cite{Parker_inj_FH61} (also called the \emph{thermal leakage}).
This scenario has been confirmed by early 1D hybrid simulations and
observations, although the contribution of the ions backscattered
from the shock interface has been recognized as well \cite{Quest88}.
Several later simulations have favored backscattered (shock reflected,
or shock dwelling) particles over those leaking from downstream \cite{Burgess12,CaprioliInj15}.
The distinction is somewhat semantic, as it is not easy to draw the
line between these groups of particles. The backscattering occurs
from a blurred and wiggling shock interface. For the thermal leakage
to occur, the thermalization of downstream ions must be fast and close
to completion. Therefore, the turbulence must be well developed to
ensure strong wave-particle interaction with an extended resonance
overlapping. These conditions are crucial for the entropy production,
e.g. \cite{ZaslavHamChaos07}, not easy to meet in simulations. 

There are many numerical limitations. To those given, e.g., in \cite{Burgess12},
we add a restricted number of wave modes and their harmonics available
for the particle thermalization downstream. Wave spectra are artificially
constrained by the size of the simulation box and its aspect ratio,
thus impeding both the turbulence development and particle scattering.
The list of limitations can be continued, but it is enough to say
that a suppression of the nonlinear Landau damping, caused by the
sparse wave spectrum, will slow down the thermalization. Recall that
it requires a beating wave resonance with particles. A scarce number
of particles in cell only contributes to this problem. We also note
here that this type of wave-particle interaction should be particularly
important for the entropy production downstream since the linear (cyclotron)
resonance with Alfven waves results primarily in pitch-angle diffusion
with almost no change in particle energy. If there is no full thermalization
downstream, the thermal leakage may also be suppressed. However, the
``numerical'' shock may compensate for the resulting deficit of
backstreaming ions by an enhanced reflection. 

The particle reflection cannot be fully understood from the hybrid
simulation as the electron kinetic effects are important \cite{Liseykina15,MSetal_IAshocks2016}.
The PIC codes are still limited in the duration of acceleration, thus
making the conclusions about the downstream thermalization mechanisms
and the origin of backstreaming particles not compelling. For example,
if all the injected protons were indeed shock reflected, as frequently
suggested by simulations, it would not be easy to explain even a few
reflected H$^{++}$, given that only a limited fraction of incoming
protons can be reflected in any event. Obviously, the both species
could not be simultaneously reflected by a stationary reflecting barrier,
be it magnetic, electrostatic, or both. If the barrier (shock overshoot)
oscillates, then the protons would still be injected more efficiently
than He ions, which does not seem to be supported by observations.
Note, however, that the He$^{++}$ leakage from downstream has been
finally obtained in simulations (Damiano Caprioli, this meeting).
This finding may be an important step towards understanding the nature
of backstreaming ions, including their composition. 

Notwithstanding the remaining uncertainties, backstreaming protons
of whatever origin are crucial for shock dissipation. They resonantly
(and partially non-resonantly) drive the MHD waves whose saturation
amplitude can be linked to the density of backstreaming protons. The
latter was obtained in \cite{m98} as a function of the wave amplitude,
and Mach number, thus making description self-consistent, at least
in its proton part. In the first-order approximation, heavier elements
are scattered by the proton-generated waves. The results of such calculations
are shown in Fig.\ref{fig:Elemental-injection-efficiencies}. Although
the proton spectra obtained in this model (not shown in the figure)
compare very well with hybrid simulations available at the time \cite{BennettElis95},
they were based on the backstreaming ions originating from the downstream
fully thermalized Rankine-Hugoniot-compliant, hydrogen-dominated plasma. 

The charge-sign dependence of injection also works against protons
\cite{m98}, but this time through the sign of the wave helicity,
they drive upstream. The mechanism is that particle orbits that spiral
along the perturbed magnetic field, have a preferred particle escape
direction that depends on the charge sign \cite{m98}. So, $\bar{p}$
and $e^{-}$, for example, would have better chances for injection
than $p$ and $e^{+}$, as they would preferentially escape upstream
through a wave generated by protons. However, both $\bar{p}$ and
$e^{+}$ need to be produced as secondaries in $pp$ collisions in
the first place and remain available for further acceleration. I will
argue in Sec.\ref{sec:ModelingPositrons} that this combination of
conditions is easier to fulfil for $e^{+}$ than for $\bar{p}$. However,
the mechanism will not be related to the wave-particle interaction.

\begin{figure}
\includegraphics[bb=20bp 100bp 550bp 540bp,  scale=0.35]{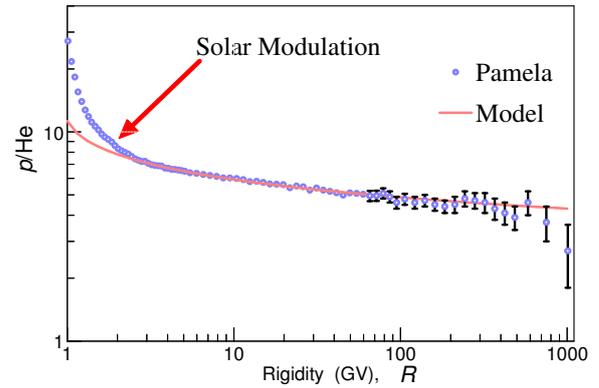}

\caption{PAMELA points (circles) and theoretical fit from \cite{MDSPamela12}
(line). The twenty highest rigidity points are shown with the error-bars
(stat.+syst.), where they seem to become significant. The difference
between the data and theory at lower rigidities is due to the heliospheric
modulation, not included in the model. This model would also fit the
later AMS-02 data equally well, as they differ from those of Pamela
only in normalization.\label{fig:PAMELA} }
\end{figure}

\selectlanguage{english}%
Using the injection rates in Fig.\ref{fig:Elemental-injection-efficiencies},
the authors of Ref.\cite{MDSPamela12} have suggested a solution to
the problem of 0.1 difference in He and proton rigidity spectra that
does not require any additional assumptions regarding SNR environments.
A single spherical SNR shock, propagating in a homogeneous medium,
\emph{must} produce this difference. The focus of their explanation
is on an initial (injection) phase of the DSA when the elemental similarity
does not apply ($\mathcal{R}<\mathcal{R}_{0}$). According to the
injection theory \cite{mv95,m98}, collisionless shocks inject relatively
more He$^{2+}$ than protons when they are stronger. Naturally, they
also produce harder spectra at that time, so the time integrated He$^{2+}$
spectrum is harder as well. The result of convolution of the relative
$p$/He injection rate with a typical SNR time history is illustrated
in Fig.\ref{fig:PAMELA}. 

\section{Modeling the $e^{+}/e^{-}$ spectrum\label{sec:ModelingPositrons}}

\paragraph*{\uline{Model Requirements and Objectives}}

Similarly to the $p$/He data discussed above, the current AMS-02
$e^{+}/e^{-}$ data, Fig.\ref{fig:PosFracAndNLDSAanalSol}, have negligible
statistical errors in an extended range between 0.5-100 GeV. The spectrum
in this range is, however, considerably more complex, thus placing
tighter constraints on the models. Therefore, a successful fit of
the data in this range lends credence to the model predictions beyond
it, including an interesting positron excess (and antiproton, for
that matter) where the data are not so exact and even nonexistent
at $E\gtrsim500$GeV. Needless to say that an incipient bump around
200-300 GeV attracted much attention from the dark matter hunters.

\paragraph*{\uline{Physics behind the Models}}

\selectlanguage{british}%
Most of the conventional scenarios for $e^{+}/e^{-}$ excess invoke
secondary positrons. They are produced by galactic CR protons in hadronic
reactions. The collisions may occur in an ambient gas near an SNR
accelerator, e.g. \cite{Fujita2009PhRvD}, elsewhere in the galaxy,
e.g., \cite{Waxman2013PhRvL,Cowsik2014ApJ}, or immediately in the
SNR shock, thus being incorporated into the DSA, e.g. \cite{Blasi2009PhRvL,Mertsch2009PhRvL}.
Some of these scenarios face the unmatched antiprotons and other secondaries
in the data, as discussed, e.g., in \cite{Mosk2002ApJ,Kachelriess11,CholisHooper2014PhRvD,YuanArxiv2017}.
Certain improvements along these lines have recently been achieved
by using sophisticated Monte Carlo $pp$ collision event generators,
e.g. \cite{Kohri2016PTEP}. However, improved cross sections of $pp$
collisions do not shed more light on the \emph{physics} of $e^{+}/e^{-}$
anomaly, particularly the minimum at 8 GeV and the U-shape form of
the spectrum. This spectrum complexity hints at richer physics than
a mere production of secondary $e^{+}$ power-law spectra from the
primary CR power-law.

Sources of positrons other than the secondaries from $pp$ collisions
have also been suggested. Apart from the pulsars \cite{Profumo2012CEJPh},
the most intriguing are dark matter (DM) related scenarios \cite{HooperDM_AMS13,CholisDMpulsars13}.
Regardless of the positron source considered, most of the models have
enough ``knobs'' to fit the data. Some SNR based approaches, e.g.,
\cite{ErlykWolf2013APh} directly use the AMS-02-measured pair of
spectral indices to reproduce the U-shape spectrum in Fig.\ref{fig:PosFracAndNLDSAanalSol}
by adjusting the weights of the source and background contributions.
The position of the spectral minimum also needs to be taken directly
from the AMS-02 data. Therefore, the physical nature of the minimum
is not convincingly interpreted by these models, thus giving no credence
to their predictions concerning the higher-energy positron excess.
It follows that the model conclusions about a possible room for DM
or lack thereof in the positron spectrum remain premature. 

\begin{figure}
\selectlanguage{english}%
\includegraphics[bb=0bp 60bp 550bp 550bp,  scale=0.35]{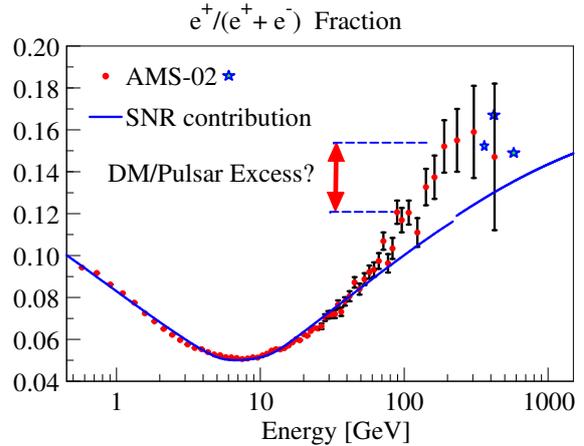}

\selectlanguage{british}%
\caption{AMS-02 data \cite{AMS02_2014}, with the error bars added only where
they are significant ($E>30$ GeV). A few recent points (stars) from
\foreignlanguage{english}{http://www.ams02.org/wp-content/uploads/2016/12/Final.pdf}
are included. Also shown is the positron fraction, obtained by solving
the nonlinear acceleration problem. The predicted saturation level
$\approx0.17$.\label{fig:PosFit}}
\end{figure}

Of course, before embarking on the quest for DM, one should look for
astrophysical interpretations of the $e^{+}/e^{-}$ and $\bar{p}/p$
anomalies more carefully. The authors of \cite{MalkPositrons2016}
give an accurate and prosaic interpretation of namely the most precisely
measured part of the positron fraction in the range 0.5-100 GeV. The
explanation is based on the following two aspects of the DSA. First,
the mechanism of positron \emph{injection} is \emph{charge-sign asymmetric,
}which contrasts with our previous discussion of the $p-$He injection
that primarily depends on the \emph{mass-to-charge} ratio. They argue
that the charge-sign selectivity of injection arises when the shock
sweeps clumps of dense molecular gas (MC).

The second aspect of the mechanism proposed in \cite{MalkPositrons2016}
concerns a nonlinear shock modification by accelerated protons (NLDSA)
which is known to make the momentum spectrum of low-energy particles
steeper and that of the high-energy particles flatter than the canonical
$p^{-4}$ spectrum produced by strong but unmodified shocks. Therefore,
there is a point $p=p_{4}$ where the index is exactly four. Assuming
most of galactic CR electrons to be accelerated in conventional shocks,
thus having $p^{-4}$ source spectra, the \emph{ratio} of the modified
positron- to unmodified electron spectrum will show the required nonmonotonic
behavior with a minimum at $p=p_{4}$. Note that, as in the case of
$p$/He spectral anomaly, it is crucial to work with the positron/electron
ratio, since the observed spectrum is automatically de-propagated
back to the source. In a customary $p^{4}f$$\left(p\right)$ normalization,
the individual positron spectrum is the same as that of the $e^{+}/e^{-}$
ratio which is, in turn, the same as the \emph{proton }spectrum, provided
that both species are accelerated to \emph{relativistic} energies
beyond their injection energies.
\selectlanguage{english}%

\paragraph{NLDSA}

\selectlanguage{british}%
The nonlinear proton spectrum is well understood since it is solvable
analytically \cite{m97a,m99,MDru01}. Excellent agreement between
the analytic and numerical solutions has been documented, e.g., in
\cite{Mosk07}, Fig.\ref{fig:PosFracAndNLDSAanalSol}. The analytic
solution places the \emph{proton} $p^{4}f$$\left(p\right)$ minimum
at $\lesssim10$ GeV/c which is encouragingly consistent with the
position of the observed $e^{+}/e^{-}$ minimum. Moreover, it depends
only weakly on the shock Mach number, $M$, proton's maximum momentum,
$p_{{\rm max}}$, and their injection rate. However, $M\gtrsim10$
and $p_{\max}/m_{p}c\gtrsim500$ are the thresholds for a bifurcation
into a strongly nonlinear (efficient acceleration) regime, shown in
Fig.\ref{fig:PosFracAndNLDSAanalSol}, with its signature minimum
around $5-10$ GeV. Understanding of this transition requires an analytic
form of the bifurcation diagram, obtained in \cite{m97b,MDru01}.
The latter cannot be inferred from most of the steady state numerical
solutions, e.g. \cite{EllisBerBar00}. When the solution is not unique,
it still converges to one particular branch and does not trace unstable
manifolds, required to understand the bifurcation phenomenon (see
below). The so-called semi-analytic treatments, e.g. \cite{Blasi02},
do not provide the bifurcation diagram (commonly called the S-curve)
either, as they do not solve the full problem in analytic form. The
significance of the S-curve is that it encompasses three co-existing
solutions in a certain range of the proton injection parameter $\nu_{1}<\nu_{{\rm inj}}<\nu_{2}$,
thus embodying hysteresis in this nonlinear system. It also provides
two critical values for injection rates $\nu_{1,2}$ as functions
of the Mach number $M$ and maximum momentum $p_{{\rm max}}.$ One
of the three solutions is unstable. The remaining two are, respectively,
an inefficient solution, which corresponds to a test-particle solution
at $\nu_{{\rm inj}}<\nu_{1}$, and the efficient solution. For $\nu_{inj}>\nu_{2}$
this solution is unique. Moreover, in the limit $M\gg1$ and $p_{{\rm max}}\gg m_{p}c$
it can be obtained \emph{analytically and exactly} \cite{m99}, which
is crucial in calculating the bifurcation diagram. 

The solution co-existence domain between the critical injection rates
$\nu_{1,2}\left(p_{{\rm max}},M\right)$ is typically very narrow,
$\nu_{2}-\nu_{1}\ll\nu_{1}$, which additionally prevents stationary
numerical solutions from accessing the hysteretic shock behavior.
By contrast, time-dependent numerical treatments should easily identify
the transition from the inefficient branch at $\nu_{{\rm inj}}<\nu_{1}$
to efficient one at $\nu_{{\rm inj}}>\nu_{2}$. Indeed, when the system
is evolved from a low $p_{\max}\lesssim m_{p}c$, at first, it sticks
with the inefficient solution. With the (slowly) growing $p_{{\rm max}}$,
the solution passes through the co-existence region, since both $\nu_{1}$
and $\nu_{2}>\nu_{1}$ decrease with $p_{\max}.$ As soon as $\nu_{2}$
becomes $\nu_{2}<\nu_{{\rm inj}}$, the only remaining system attractor
(stationary solution under fixed $p_{{\rm max}}$) is the efficient
solution, to which the numerical time-dependent solution will quickly
converge. This type of dynamics was likely seen in time-dependent
kinetic simulations \cite{KangJones2002}. It occurred there in form
of a series of sudden jumps to higher shock pre-compressions, manifesting
the transitions from inefficient to efficient solutions when the former
cease to exist ($\nu_{{\rm inj}}>\nu_{2}$). This pattern of the shock
evolution was predicted in \cite{m97b,MDru01}, including back transitions,
though the authors of \cite{KangJones2002} attributed it to a likely
numerical artifact. The incipient back transitions make the time-dependent
evolution oscillatory in character, as seen in the simulations. The
parameters of these limit-cycle oscillations (related to a Hopf bifurcation)
depend on how other control parameters (primarily $\nu_{{\rm inj}})$
respond to the changing shock compression during the forward or back
transitions. This feedback loop is self-consistently included in the
above simulations. Therefore, the limit-cycle type oscillations obtained
in \cite{KangJones2002} are likely to be genuine and worth further
study. It will help to understand the NLDSA dynamically, thus shedding
new light on many shock-related astrophysical sources where a short-time
variability is observed. 

It should be noted, however, that a significant turbulent heating
upstream is likely to straighten out the S-curve, thus making only
one solution possible \cite{MDru01}. A similar effect of heating
on the shock solutions occurs in some CR fluid models, e.g. \cite{MostafaviZank2016AIPC}.
The turbulent heating impacts the NLDSA in many ways but it is difficult
to quantify, as the turbulence in the nonlinear shock precursor is
still not well understood \cite{MDS10PPCF}.
\selectlanguage{english}%

\paragraph{Interaction with MCs}

\selectlanguage{british}%
Due to the sub-shock weakening in the NLDSA, MCs survive the ionizing
radiation. Moreover, shock-accelerated CR protons illuminate the MC
well before the subshock encounter. These CRs generate positrons and
other secondaries in the MC interior by $pp$ collisions. They also
charge the MC \emph{positively} which creates a charge-sign asymmetry
for the subsequent particle injection into the DSA via the following
simple mechanism \cite{MalkPositrons2016}. 

Because of a positive electrostatic potential in an MC, built by penetrating
shock-accelerated protons, low-energy positively charged particles
will be expelled from the MC, while negatively charged particles stay
inside. Negatively-charged low-energy secondaries produced at the
periphery of the MC by relatively low-energy but abundant CR protons
will be sucked into the MC. This phenomenon is particularly consistent
with the $\bar{p}/p$ decline towards lower energies, Fig.\ref{fig:AntiprotonsAMS}.
However, for kinematic reasons of their generation and large mass
ratio $m_{p}/m_{e}$, antiprotons will be absorbed by the MC much
less efficiently than electrons. More importantly for the sequel,
the same electric field expels the positively charged positrons from
the MCs very efficiently. This is also consistent with an increase
of the $e^{+}/e^{-}$ ratio, towards lower energies Fig.\ref{fig:PosFracAndNLDSAanalSol}. 

The charging of MC by accelerated protons is a complex phenomenon
\cite{MalkPositrons2016} and merits a short discussion. While fully
ionized plasmas are intolerant to external charges and immediately
restore charge neutrality, sufficiently large and dense MCs respond
differently. Due to a high rigidity of CRs, their density in the MC
interior increases almost simultaneously with their density outside,
when a CR-loaded shock approaches the MC. However, by contrast with
a strongly ionized exterior, where the plasma resistivity is negligible,
the electron-ion collisions inside the MC provide sufficient resistivity
to neutralizing electric currents. Therefore, a strong macroscopic
electric field will build up in response to the CR penetration, to
neutralize the CR charge. This field expels the secondary positrons
most efficiently as the lightest positively charged species \textendash{}
although it also shields the MC from low-energy CR protons.

The mechanism outlined above implies that negatively charged primaries
and secondaries have much better chances to stay in an MC than positively
charged particles. When the subshock eventually reaches the MC, the
subshock engulfs it, e.g., \cite{Inoue12,Draine2010BookISM}. What
was in the MC interior, is transferred downstream, largely unprocessed
by the subshock. Therefore, the negatively charged particles mostly
evade acceleration which explains why there is no $\bar{p}/p$ excess
similar to that of $e^{+}/e^{-}$ at high energies, cf. Figs.\ref{fig:PosFit}
and \ref{fig:AntiprotonsAMS}. Also, antiprotons are strongly depleted
towards lower energies, since they are likely to be trapped by the
MC electrostatic potential, as discussed above. Antiprotons born in
$pp$ collisions with higher energies (typically a few GeV) can be
retained in the MC only if their are deep inside. Otherwise, they
are accelerated just the way the protons are, which explains the flat
$\bar{p}/p$ ratio at higher energies. 

\section{Conclusions and Outlook}

The common belief that \foreignlanguage{english}{the CR rigidity spectrum
is a power-law with only a few distinct structures (e.g., knee and
ankle), the same for all primary elements, is being rapidly abandoned.
New data have challenged the idea of acceleration scalability on the
entire rigidity range of a given accelerator, such as an SNR. }

\selectlanguage{english}%
Multiple sources (SNRs) with adjustable spectra or sources with poorly
known spectra (DM, pulsars, superbubbles with uncertain environments)
were almost universally proposed to accommodate the new spectral features.
By contrast, the gist of this paper was that a \emph{single-SNR} CR
acceleration allows for much more certain predictions about rigidity
spectra of different elements. Moreover, some SNRs are accessible
to direct observations across a broad emission spectrum. The CR production
in such sources, including the CR interaction with adjacent MCs \cite{MDS_11NatCo},
is therefore reasonably well understood which provides further insights
into the nature of CR anomalies discussed in this paper. It was demonstrated
that it is possible to fit a high-fidelity part of the data, provided
that the following elements of the DSA physics are addressed 
\begin{itemize}
\item self-consistent particle injection, depending on mass to charge ratio
and shock Mach number
\item clumps of a dense gas present in SNR shock environments
\item backreaction of accelerated protons on the shock and all particle
injection and acceleration
\end{itemize}
It is clear that the proposed single-source mechanisms equally apply
to an ensemble of similar SNRs without introducing additional free
parameters. A possible exception is an upturn in $p$-He rigidity
spectra near 300 GV, not discussed in this paper. Its most obvious
explanation is that a steeper, low-energy part of the spectrum is
produced by a nearby SNR. At rigidities $>300$ GV, it ``sinks''
into a slightly harder spectrum, which is either created by a stronger
(younger) but more remote object, or else represents galactic background.
The relative normalization of the two contributions needs then to
be adjusted to place the kink in the right place in the spectrum.
The upturn, though, might also be explained by a CR self-confinement
effect around the source via CR-emitted Alfven waves \cite{MetalEsc13}.
It is not clear, however, if the latter scenario can be elaborated
with a lesser number of free parameters than the two-source model.
If the two-source scenario is at work, the following conclusion, relevant
to the present study, can be drawn from the 300 GV upturn. Because
the $p$/He spectrum does not show any change in its slope at this
rigidity, the mechanism of $p$ and He injection in both sources is
likely to be universal, as argued in this paper. 

Further insights into the physics of injection give much improved
numerical simulations. However, as they are becoming more realistic,
their physical understanding is often a challenge. The present day
simulations can be considered as numerical experiments, not much easier
to interpret than real observations. Understanding their limitations
by benchmarking between different software suites and comparison with
simplified physical models is a promising avenue to understanding
rapidly improving observations. 

To conclude, I briefly dwell on the prospects for resolving the remaining
mismatch with the positron fraction in the 200-400 GeV range, Fig.\ref{fig:PosFit}.
Assuming that the data, notwithstanding error bars in this range,
are statistically significant, one may suggest a few explanations
of this deviation. The most exciting one is that the limited-energy-range
excess over the SNR positron production manifests a DM annihilation
or decay. Ironically, the model described in this paper that was incited
to account for the data astrophysically facilitates the DM scenario
by providing an SNR ``background''. It significantly eases the requirements
for the positron production by DM particles. Not going deeper into
DM scenarios, several conventional ways to explain the deviation,
may also be suggested \cite{MalkPositrons2016}. Further elaborations
on them may or may not close the remaining gap between the SNR positron
production and the AMS-02 data. For now, the possibility of DM or
pulsar contribution to the positron excess cannot be ruled out, so
the prospects for the future of the CR research are exciting.

\section*{Acknoledgements}

I'm grateful to the organizers for inviting me to speak at this meeting.
I also thank Roger Blandford, Igor Moskalenko, and Paul Simeon for
useful discussions we have had afterward in Palo Alto, California.

This work was supported by the NASA Astrophysics Theory Program under
Grant No. NNX14AH36G.


\bibliographystyle{elsarticle-numCutAuthList}
\bibliography{Malkov.bbl}
\end{document}